# On the Possibility of Singlet Fission in Crystalline Quaterrylene

*Xiaopeng Wang,[1] Xingyu Liu,[1] Cameron Cook,[2] Bohdan Schatschneider,[2] and Noa Marom[1,3,4,*]*

1. Department of Materials Science and Engineering, Carnegie Mellon University, Pittsburgh, PA 15213, USA
2. Department of Chemistry and Biochemistry, California State Polytechnic University at Pomona, Pomona, CA 91768, USA
3. Department of Chemistry, Carnegie Mellon University, Pittsburgh, PA 15213, USA
4. Department of Physics, Carnegie Mellon University, Pittsburgh, PA 15213, USA

**Corresponding Author:** nmarom@andrew.cmu.edu

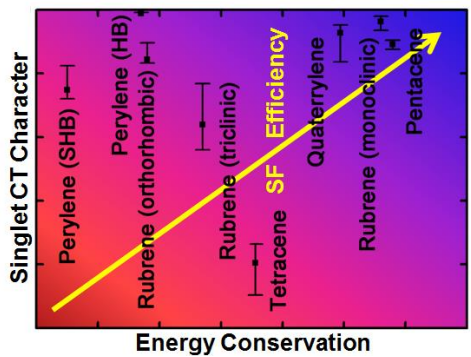

## Abstract

Singlet fission (SF), the spontaneous down-conversion of a singlet exciton into two triplet excitons residing on neighboring molecules, is a promising route to improving organic photovoltaic (OPV) device efficiencies by harvesting two charge carriers from one photon. However, only a few materials have been discovered that exhibit intermolecular SF in the solid state, most of which are acene derivatives. Recently, there has been a growing interest in rylenes as potential SF materials. We use many-body perturbation theory in the *GW* approximation and the Bethe-Salpeter equation (BSE) to investigate the possibility of intermolecular SF in crystalline perylene and quaterrylene. A new method is presented for determining the percent charge transfer character (%CT) of an exciton wave-function from double-Bader analysis. This enables relating exciton probability distributions to crystal packing. Based on comparison to known and predicted SF materials with respect to the energy conservation criterion ($E_S - 2E_T$) and %CT, crystalline quaterrylene is a promising candidate for intermolecular SF. Furthermore, quaterrylene is attractive for OPV applications thanks to its high stability and narrow optical gap. Perylene is not expected to exhibit SF, however it is a promising candidate for harvesting sub-gap photons by triplet-triplet annihilation.

# Introduction

Many applications of organic solids, ranging from sensors to photovoltaics,[1-3] involve photon absorption and charge generation. Organic chromophores used in these devices have the advantages of being low-cost, lightweight, and having chemically tunable properties. Previously, the Shockley-Queisser limit,[4] stating that one photon only creates one electron-hole pair, placed an upper bound on the efficiency of organic photovoltaics (OPV). However, this limit may be surpassed by singlet fission (SF),[5-13] a collective many-body quantum mechanical process whereby one photogenerated singlet exciton splits into two triplet excitons located on different molecules, generating two carriers for harvesting. SF was first observed in crystalline anthracene in 1965.[14] Recently, there has been a surge of renewed interest in SF following the realization that it may improve the theoretical solar conversion efficiency limit from 33% to 47%.[4,15]

The number of materials reported experimentally to exhibit SF to date is very small,[10,16,17] considering the infinitely vast chemical compound space. For the occurrence of SF to be thermodynamically favorable, the singlet excitation energy ($E_S$) should be larger, or at least not much smaller than twice the triplet excitation energy ($E_T$). If this condition is met, SF will be exoergic or not significantly endoergic.[5,18] This energy conservation criterion excludes most organic chromophores. Furthermore, for SF to be observed experimentally, its rate should be faster than all competing processes, such as fluorescence, intramolecular relaxation processes, and in particular triplet-triplet annihilation (TTA), which is the reverse process of SF.[5,10,16,19] In this respect, meeting the energy conservation criterion also helps suppress TTA. Long exciton lifetimes may be crucial to preventing singlets from decaying via fluorescence and enhancing triplet harvesting. It has been postulated that a singlet exciton with charge-transfer (CT) character (i.e., with the electron probability density concentrated away from the hole position) would be favorable for intermolecular SF because it could couple more easily with triplet excitons localized on neighboring molecules, while suppressing electron-hole recombination.[5-7,20-22] For practical photovoltaic device applications, there are additional requirements of SF materials, such as optical absorption of a broad range of the solar spectrum and long-term chemical stability under operating conditions. These stringent requirements leave much to be desired and call for exploration of new SF candidates.

Oligoacenes[23-34] and their derivatives[35,36] stand out among organic chromophores and have been pivotal in fundamental research of SF. Both tetracene[37-41] and pentacene[42-44] are top performers, exhibiting fast SF and near 200% triplet yield.[10,37,42] To elucidate the success of the acene family, excitonic properties have been investigated both experimentally and theoretically, especially in crystalline pentacene, which satisfies the energy conservation criterion, unlike most other organic crystals. In a recent computational study,[43] the wave-function of the lowest energy singlet exciton in pentacene was found to have 94% CT character, which may contribute to accelerating the SF process. SF has also been observed in several acene derivatives, such as bis(triisopropylsilylethynyl) pentacene (TIPS-pentacene)[45,46] and orthorhombic rubrene

(which may be considered as functionalized tetracene).[47-49] The two meta-stable monoclinic and triclinic polymorphs of rubrene, which have not been widely studied experimentally, have been predicted theoretically to exhibit more efficient SF than the orthorhombic form.[49] Inspired by the success of oligoacenes and their derivatives, we turn to other conjugated hydrocarbons in search of new promising candidates for SF.

Oligorylenes[50-55] are another family of polycyclic aromatic hydrocarbons (PAHs) with unique thermodynamic, electronic, and excitonic properties, including SF.[56,57] The smallest member of this family, perylene, shown in Figure 1, has been reported to undergo SF from super-excited singlet states, as well as TTA.[58] Recently, SF has been measured experimentally in crystalline perylene derivatives.[59-65] In particular, a triplet yield of 140±20% in 180±10 ps was reported for perylenediimide thin films.[61] The second member, terrylene, may be superior to perylene, based on the experimental observation of endoergic SF by only 70 meV and close to 200% triplet yield in two tert-butyl-substituted terrylenes.[66] In both perylene and terrylene derivatives[67,68] SF quantum yield may be further enhanced by manipulating the crystal packing,[59] similar to rubrene.[49,69] A theoretical study of gas-phase oligorylenes, using time-dependent density functional theory (TDDFT), has suggested that larger molecules of this family are more likely to meet the energy conservation criterion,[56] however this study did not consider intermolecular SF in the solid state. In molecular crystals the molecular orbitals evolve into dispersed bands and excitons may be delocalized over several molecules.[21,43] This may facilitate the coupling between the singlet exciton and the triplet excitons of neighboring molecules, contributing to efficient intermolecular SF.[6,7,20,22]

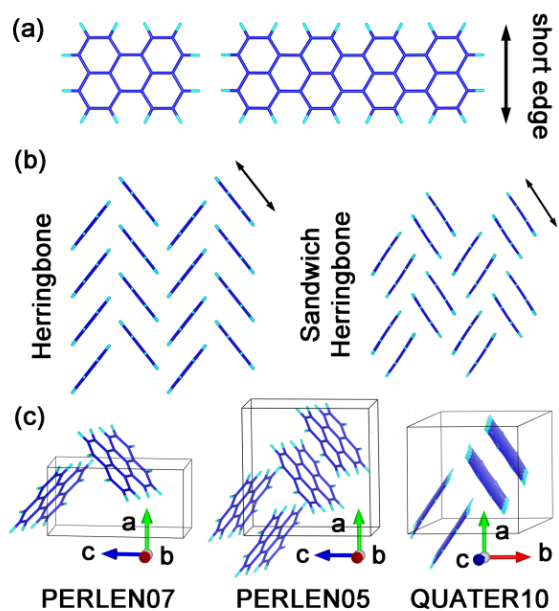

**Figure 1**. (a) Perylene (left) and quaterrylene (right) single molecules, (b) the herringbone (HB) and sandwich-herringbone (SHB) packing of one molecular layer in the two perylene polymorphs viewed along the long edge of molecules, and (c) the unit cells of HB perylene, SHB perylene, and SHB quaterrylene with their respective CSD reference codes. C and H atoms are shown in dark and light blue, respectively.

We investigate crystalline perylene[70-72] and quaterrylene,[50,73,74] shown in Figure 1, as candidates for intermolecular SF (no crystal structure has been reported for terrylene, therefore it is not considered here). Quaterrylene crystallizes in a sandwich-herringbone (SHB) packing motif[74,75] with four molecules per unit cell. Perylene has two known polymorphs. The $\alpha$ form has a SHB structure, similar to quaterrylene. The $\beta$ form has a herringbone (HB) structure[75] with two molecules per unit cell, similar to tetracene and pentacene. To describe the excited state properties of perylene and quaterrylene we use a beyond-DFT approach, based on Green's function many-body perturbation theory (MBPT). The *GW* approximation,[76-79] where *G* is the one-particle Green's function and *W* is the screened coulomb interaction, is used to calculate properties derived from charged excitations, such as fundamental gaps and band structures. *GW* accounts for the dynamical correlation (screening) effects that cause the renormalization of quasiparticle energies in response to particle addition/removal. *GW* quasiparticle energies are then used to construct the two-particle Green's function, which describes coupled electron-hole excitations, and then solve the Bethe-Salpeter equation (BSE)[78,80] to obtain properties derived from neutral excitations, including the optical absorption spectra, singlet and triplet energies, and exciton wave functions. A new method of double-Bader analysis is introduced to evaluate the degree of CT character of exciton wave functions produced by BSE calculations. Perylene and quaterrylene are then compared to pentacene, tetracene, and the three polymorphs of rubrene with respect to a two-dimensional descriptor based on the energy conservation criterion ($E_S$-$2E_T$) and the degree of singlet exciton CT character (%CT). The comparison between eight molecular crystals reveals trends across chemical families and packing motifs (the computational cost of *GW*+BSE calculations prohibits screening a large number of materials). We find crystalline quaterrylene to be a very promising candidate for intermolecular SF.

## Methods

### Double-Bader Analysis of Exciton Character

Bader analysis[81,82] is a widely used partitioning scheme applied to charge densities, $\rho(r_e)$, with one electron position variable, $r_e$, in the format of volumetric data on a discrete three-dimensional spatial grid. For each atom, a Bader volume is defined,[82,83] which contains a single electron density maximum and is separated from other volumes by a zero flux surface of the gradients of the electron density. Once the hole position, $r_h$, is fixed, the two-particle exciton wave-function from a BSE calculation, $|\Psi(r_e, r_h)|^2$, becomes an electron charge density with only one electron position variable, $|\Psi(r_e, fixed\ r_h)|^2$, and thus may also be treated with Bader analysis. Based on this, we propose a new method for calculating the degree of CT or Frenkel (*i.e.,* the electron and hole are located on the same molecule) exciton character by a "double-Bader" analysis, performing nested sums over electron and hole positions. As illustrated schematically in Figure 2 we define the probability of finding a hole and an electron, respectively, on molecules A and B in the supercell as:

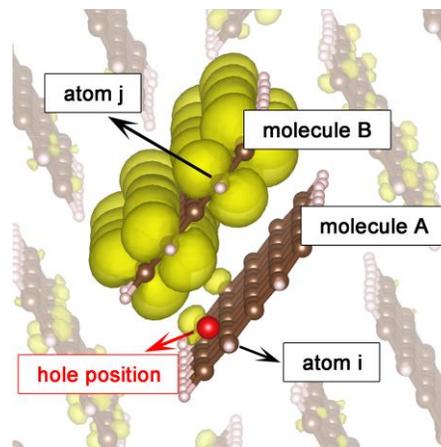

**Figure 2.** Schematic illustration of the double-Bader analysis. A hole, denoted by a red dot, is placed in the Bader volume of atom *i* of molecule A. The electron probability density corresponding to this hole position is shown in yellow. The inner sum in equation (1) runs over all atoms *j* of molecule B. The outer sum is over all symmetry-inequivalent atoms of molecule A. The middle sum is over the number of hole positions sampled in the Bader volume of each atom *i* of molecule A.

$$P_{e@molecule\ B}^{h@molecule\ A} = \sum_{atom\ i\ \in\ A} W_i \sum_{r_h\ \in\ atom\ i} \frac{1}{N_{r_h\ \in\ atom\ i}} \sum_{atom\ j\ \in\ B} P_{e@atom\ j}^{h@fixed\ r_h\ \in\ atom\ i} \quad (1)$$

$W_i$ is a weight corresponding to the relative probability of finding a hole in the Bader volume of atom $i$ in molecule A and $P_{e@atom\ j}^{h@fixed\ r_h\ \in\ atom\ i}$ is the probability of finding the electron in the Bader volume of atom $j$ in molecule B when the hole is located in the Bader volume of atom $i$, both of which are calculated using Bader analysis.[81,82] $N_{r_h\ \in\ atom\ i}$ is the number of hole positions sampled in the Bader volume of atom $i$.

The lowest-energy singlet excitation in molecular crystals typically corresponds to transitions from the top valence bands to the bottom conduction bands, derived from the single molecule highest occupied molecular orbital (HOMO) and lowest unoccupied molecular orbital (LUMO), respectively (for quaterrylene, for example, four bands are derived from the four molecules in the unit cell). Therefore, the electron charge density distribution of the single molecule HOMO is a reasonable estimate for the hole probability. We perform Bader analysis of the HOMO electron density obtained from a DFT calculation of a gas-phase molecule to determine the weight factors, $W_i$, corresponding to the relative probability of a hole positioned at atom $i$. The probability of finding an electron on atom $j$, $P_{e@atom\ j}^{h@fixed\ r_h\ \in\ atom\ i}$, is then obtained via Bader analysis of the BSE exciton wave-function, $|\Psi(r_e, fixed\ r_h)|^2$. $W_i$ corresponds to the hole probability in the whole Bader volume centered at atom $i$, while $P_{e@atom\ j}^{h@fixed\ r_h\ \in\ atom\ i}$ is calculated with $r_h$ placed at a fixed position within this volume. Therefore, $r_h$ should be sampled at enough positions, $N_{r_h\ \in\ atom\ i}$, within the Bader volume of atom $i$ to calculate the averaged $P_{e@atom\ j}^{h@fixed\ r_h\ \in\ atom\ i}$ for each $W_i$. In the case of quaterrylene, changing the hole position from 0.2 Å above an atom to 0.4, 0.6, 0.8 and 1.0 Å resulted in negligible change of the exciton wave-function (see SI). Therefore, the hole position could be sampled at only one position, e.g. 0.8 Å above atom $i$ in molecule A, to evaluate $P_{e@atom\ j}^{h@fixed\ r_h\ \in\ atom\ i}$. In crystalline quaterrylene there is one cofacial neighbor on one side of each molecule and two herringbone neighbors on the other side. Therefore, two positions were sampled, one on each side, for each atom $i$. The middle sum in equation (1) is for $r_h$ sampling within the Bader volume of atom $i$. The outer and inner sums are over hole and electron probabilities on atoms $i$ and $j$ belonging to molecules A and B, respectively. The percentage of CT character (%CT) is defined as the probability for the electron and hole not to be located on the same molecule:[43]

$$\%CT = \left(1 - \frac{1}{Z}\sum_Z P_{e@molecule\ A}^{h@molecule\ A}\right) \cdot 100\% \quad (2)$$

where the degree of Frenkel character is averaged over $Z$ inequivalent molecules in the unit cell.

Recently, Sharifzadeh et al. proposed a different approach to quantitatively evaluate exciton character,[43] which involves sampling two hole positions for each C atom in the primitive cell. We only consider inequivalent atoms on inequivalent molecules in the primitive cell, which significantly reduces the number of hole positions sampled. In the double-Bader approach, the dependence of the exciton character on the hole position in the unit cell is captured by the intermediate results with $r_h$ fixed at different positions.

Error bars for %CT are then estimated from the maximal and minimal values obtained with $r_h$ fixed on different atoms in inequivalent molecules A:

$$\%CT_{atom\ i\ \in\ A} = \left(1 - P_{e@molecule\ A}^{h@fixed\ r_h\ \in\ atom\ i}\right) \cdot 100\% \qquad (3)$$

Furthermore, for a molecule possessing special chemical groups (e.g., electron donating or withdrawing groups), hole and electron probabilities may be selectively evaluated on these special groups in the outer and inner sums of equation (1), respectively, to quantify their effect on the exciton character. Although perylene and quaterrylene do not have any functional groups, this may be a useful feature of our method for other systems. Thus, the double-Bader method provides a quantitative way to analyze the exciton wave-functions. A detailed step-by-step example of the double-Bader analysis for quaterrylene is provided in the supplemental information (SI). Similar analyses were carried out for the lowest energy singlet excitons of HB perylene, SHB perylene, pentacene, tetracene, and the three polymorphs of rubrene. For pentacene we obtain 97% CT, in good agreement with the result of 94% CT, reported by Sharifzadeh *et al.*[43]

**Computational Details**
Initial geometries of crystalline HB perylene, SHB perylene, quaterrylene, and tetracene were obtained from the lowest temperature data sets available in the Cambridge Structural Database (CSD) with reference codes of PERLEN07,[71] PERLEN05,[72] QUATER10[50] and TETCEN01.[84] Geometry optimization was performed with the CASTEP code[85] using Perdew, Burke, and Ernzerhof (PBE)[86,87] coupled to the Tkatchenko-Scheffler (TS)[88] pairwise dispersion method. Norm-conserving pseudopotentials were utilized for C and H atoms. The planewave basis set cutoff was 750 eV and a Monkhorst-Pack k-grid with spacing of about 0.07 Å$^{-1}$ was used. The convergence criteria for total energy, maximum force, maximum stress, and maximum displacement were 5×10$^{-6}$ eV/atom, 0.01 eV/Å, 0.02 GPa, 5×10$^{-4}$ Å$^{-1}$, respectively. The relaxed structures are in good agreement with experiments, as shown in the SI.

The *GW* approximation and Bethe-Salpeter equation (BSE), as implemented in the BerkeleyGW code[89] were used to calculate the electronic and optical properties of crystalline HB perylene, SHB perylene, quaterrylene, and tetracene, similar to our previous work.[49] First, DFT eigenvectors and eigenvalues were generated with Quantum Espresso,[90] using the PBE exchange-correlation functional (*GW* calculations for perylene and quaterrylene molecules based on different starting points are provided in the SI). Trouiller-Martins[91] norm-conserving pseudopotentials with the $2s^2$ and $2p^2$ states considered as valence for C were generated with FHI98PP.[92] The DFT calculation was performed with k-grids of 4×4×2, 2×2×2, 2×2×2 and 4×4×2 for HB perylene, SHB perylene, quaterrylene, and tetracene, respectively. The kinetic energy cutoff was 50 Ry. Second, non-self-consistent $G_0W_0$ was employed to compute quasiparticle band structures. This is denoted as $G_0W_0$@PBE. The dielectric function and self-energy operator were constructed by summing over 550 unoccupied bands, for HB perylene, SHB perylene, and quaterrylene, and 556 unoccupied bands for tetracene. The static remainder correction[93] was applied to accelerate convergence with respect to the number of unoccupied states. An energy cutoff of 10 Ry was adopted to truncate the sums used for the calculation of the polarizability. Band structures were calculated along the high symmetry directions suggested in Ref. 94.

Lastly, the optical excitation properties were obtained by solving the BSE within the Tamm-Dancoff approximation (TDA).[89] This is denoted as $G_0W_0$+BSE@PBE. For the BSE calculation, denser k-grids of 5×5×4, 4×4×4, 4×4×4 and 8×8×4 were used for HB perylene, SHB perylene, quaterrylene, and tetracene, respectively. 24 valence bands and 24 conduction bands were considered. The polarization of light was directed along the three crystal axes. The experimental absorption spectrum of quaterrylene was measured along an unknown direction in Ref. 95. Therefore, the total absorption spectrum was simulated by combining the calculated spectra along three directions with weights providing the best fit to experiment. The lowest energy singlet and triplet exciton wave-functions were calculated to describe the possibility of finding electron in 10×3×3, 4×4×4, 4×4×4 and 8×8×4 extended HB perylene, SHB perylene, quaterrylene, and tetracene supercells, respectively. The hole position was fixed at a high hole probability site, determined based on the DFT HOMO density. The HB perylene supercell was particularly extended along the *a* direction because the singlet exciton wave-function is significantly extended along the molecular cofacial stacking direction, requiring a larger distance to converge. The results for tetracene are provided in the SI.

## Results and discussion

### Electronic and Optical Properties of Perylene and Quaterrylene

In molecular crystals the discrete energy levels of gas phase molecules turn into dispersed energy bands. The frontier molecular orbitals of the chromophores considered here are derived from the carbon π-electron density and are orientated above and below the plane of the molecule. Therefore, the band dispersion, determined by the electronic coupling between molecules in the crystal, may be related to the presence of cofacial intermolecular interactions. These cofacial interactions may be visualized *via* the deconvoluted C⋯C Hirshfeld surface, as discussed in the SI. Significant cofacial intermolecular interaction leads to larger intermolecular overlap integrals, producing more dispersed bands.

Figure 3 shows the $G_0W_0$@PBE quasiparticle band structures of crystalline perylene and quaterrylene. The main features

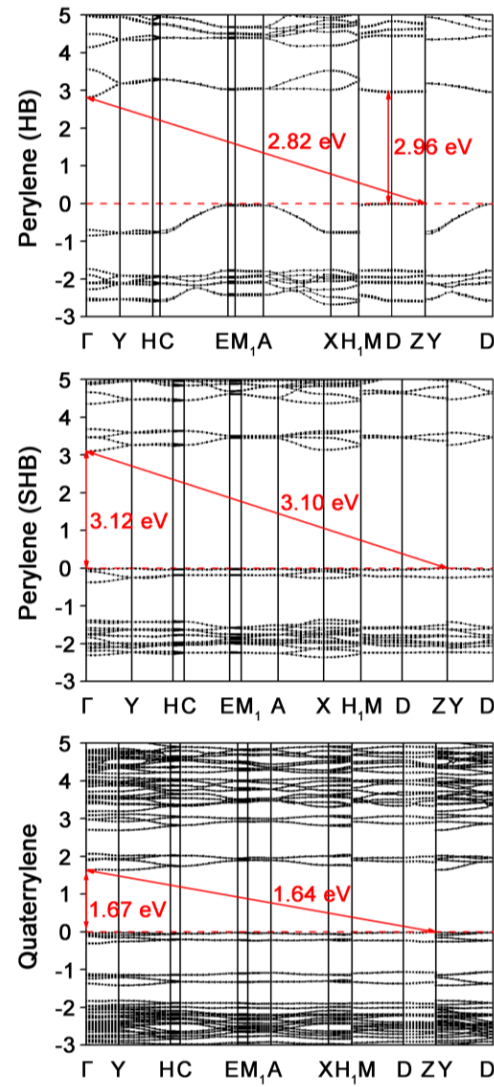

**Figure 3.** $G_0W_0$@PBE quasiparticle band structures of perylene and quaterrylene.

are summarized in Table 1 and compared to pentacene, tetracene, and the three polymorphs of rubrene. The top valence bands and bottom conduction bands (two bands for HB perylene and four for SHB perylene and quaterrylene) are derived from the single molecule HOMOs and LUMOs, respectively, and are energetically separated from other valence and conduction bands by 0.5 to 1.0 eV in all three band structures. This is consistent with the isolated HOMO peaks in the gas phase spectra, shown in the SI. The dispersion of the HOMO-derived bands in HB perylene is 0.9 eV, much larger than 0.4 and 0.3 eV, respectively, in SHB perylene and quaterrylene. This may be attributed to the different packing motifs, shown in Figure 1 (b).

In the HB polymorph of perylene the molecules are stacked, such that each molecule has two cofacial neighbors along the *a* direction, leading to continuous intermolecular coupling along the stacking direction and a large band dispersion. In the SHB structure each molecule has only one cofacial neighbor, leading to less continuous C···C interactions throughout the matrix, and therefore weaker intermolecular coupling. Pentacene and tetracene exhibit HB packing, similar to perylene, however their top valence band dispersions are only 0.5 and 0.4 eV, respectively (see the SI of this work and of Ref. 49). This may be attributed to the smaller cofacial interaction between molecules in the stacking direction as evident from the diminutive amount of C···C contacts produced from Hirshfeld surface analysis (See SI).

The fundamental band gaps of HB perylene, SHB perylene, and quaterrylene are 2.82 eV, 3.10 eV, and 1.64 eV, respectively, significantly smaller than the gas phase HOMO-LUMO gaps of perylene, 5.13 eV, and quaterrylene, 3.28 eV (see SI). This is due to polarization induced gap narrowing in extended systems.[96-98] Notably, quaterrylene has a smaller gap than the known SF materials listed in Table 1 and most other PAHs.[75] A narrow gap is desirable for solar cell applications because it enables absorption of a broader spectral range.

**Table 1**. Summary of computed excitonic properties, including fundamental gap ($E_g$), lowest-energy singlet and triplet excitation energies ($E_{S/T}$), the energy conservation condition for SF ($E_S$-2$E_T$), and %CT for the lowest lying singlet excitons.

| (eV) | $E_g$ | $E_S$ | $E_S^{exp}$ | $E_T$ | $E_T^{exp}$ | $E_S$-2$E_T$ | %CT |
|---|---|---|---|---|---|---|---|
| Perylene (HB) | 2.82 | 2.46 | 2.65[70] | 1.55 | | -0.64 | 99.7% |
| Perylene (SHB) | 3.10 | 2.61 | 2.65[70] | 1.68 | | -0.75 | 93.7% |
| Quaterrylene | 1.64 | 1.33 | 1.48[95] | 0.82 | | -0.31 | 98.2% |
| Tetracene | 2.70 | 2.24 | 2.38[99,100] | 1.34 | 1.25[28,106] | -0.44 | 80.1% |
| Pentacene | 2.25[49] | 1.72 | 1.85[101] | 0.97 | 0.86[28,106,107] | -0.22 | 97.3% |
| Rubrene (monoclinic) | 2.96[49] | 2.56 | 2.36[102] | 1.40 | | -0.24 | 99.1% |
| Rubrene (orthorhombic) | 2.69[49] | 2.28 | 2.31-2.32[103-105] | 1.45 | 1.14[47,108-110] | -0.62 | 96.1% |
| Rubrene (triclinic) | 2.68[49] | 2.25 | 2.31-2.35[102] | 1.39 | | -0.53 | 91.0% |

Figure 4 shows the $G_0W_0$+BSE@PBE optical absorption spectra of perylene and quaterrylene, compared to experimental data. The main features are summarized in Table 1 and compared to pentacene, tetracene, and the three polymorphs of rubrene. For both perylene polymorphs, optical absorption spectra were measured[70] and calculated for light polarized along the *a* and *c* directions. The single crystal absorption spectrum of quaterrylene was measured along an unknown direction.[95] Therefore, spectra computed for light polarized along the three crystal axes were combined with weights providing the best fit to experiment. The calculated optical gaps of HB perylene, SHB perylene, and quaterrylene, which correspond to the lowest energy singlet excitation, are underestimated by 0.19 eV, 0.04 eV, and 0.15 eV, respectively. For HB perylene the computed spectra agree well with experiments, except for an underestimation of about 0.3 eV in the positions of the peaks at 4.8 and 4.7 eV in the *a* and *c* directions, respectively. The computed spectra of SHB perylene agree qualitatively with experiments, however the sharp peak at 2.7 eV along the *a* direction is seen as a broad spectrum in experiments, and the absorption intensity in the region around 4.0 eV is overestimated along the *a* direction, compared to the *c* direction. For quaterrylene, the observed peaks may be attributed mainly to the spectra for light polarized along the *a* and *b* directions, while the *c* direction has a very minor contribution. The two peaks centered at 2.5 and 3.0 eV are reproduced well in the simulated total spectrum, and can be attributed primarily to the *b* direction. Additionally, the relative intensity of the strong peak at 1.4 eV compared to the aforementioned two peaks is in agreement with experiment. However, this strong peak at 1.4 eV, mainly contributed by the *a* direction spectrum, is shifted to a lower energy compared with the experimental peak position of 1.7 eV. The solar conversion efficiency of a device is limited by several unavoidable losses,[4] one of which is that photons with energy below the optical gap cannot be absorbed. Overall, quaterrylene exhibits a broader absorption spectrum than the two polymorphs of perylene. Among the eight materials listed in Table 1, crystalline quaterrylene has the smallest optical gap leading to the broadest absorption region.

Differences between the simulated and measured optical spectra (in particular for SHB perylene) may be the result of an accumulation of errors from approximations used in all three steps of the calculation, starting from DFT, through $G_0W_0$, to BSE. Sources of errors in *GW* and BSE calculations are discussed extensively in e.g., Refs. 79, 111-118. Excitonic properties may be sensitive to differences in the relaxed geometry obtained with different DFT methods.[28,119] PBE+TS used here, has provided reliable predictions for the structure of PAH crystals,[27,75,120-122] however somewhat larger deviations from experiment are obtained for HB perylene, as discussed in the SI. Errors in the $G_0W_0$ step may stem from the $G_0W_0$ approximation itself (i.e., neglecting the vertex, lack of self-consistency, and the diagonal approximation),[123-128] numerical settings, pseudopotentials,[129,130] the mean-field starting point (e.g., self-interaction errors in DFT functionals),[111,131-133] and approximations used in the self-energy evaluation, such as the Hybertsen-Louie generalized plasmon-pole model.[89] *GW*+BSE only considers particle-hole interactions and cannot describe states with multi-exciton character[134,135] (it has been postulated based on multi-reference calculations that one of the low-lying excitons of the quaterrylene molecule in the gas phase may have a fraction of double-excitation character,[136,137] however such calculations cannot be performed for molecular crystals with periodic boundary conditions). Furthermore, only direct transitions are considered in the BSE calculation. Perylene, quaterrylene, and tetracene have indirect band gaps, which are slightly smaller than their direct gaps. Contributions from indirect transitions could slightly lower the optical absorption threshold. The optical spectrum is sensitive to numerical settings, such as *k*-point sampling, which must be carefully converged.[49] The TDA, which amounts to neglecting the resonant anti-resonant coupling term, may lead to errors in both peak position and intensity.[29,112-114] Finally, dynamic and thermal effects, such as geometry relaxation in the excited state and coupling to phonons, are not considered here. Despite its limitations, *GW*+BSE is the present state-of-the-art method for calculating the excitonic properties of large periodic systems with a few hundred atoms in the unit cell.

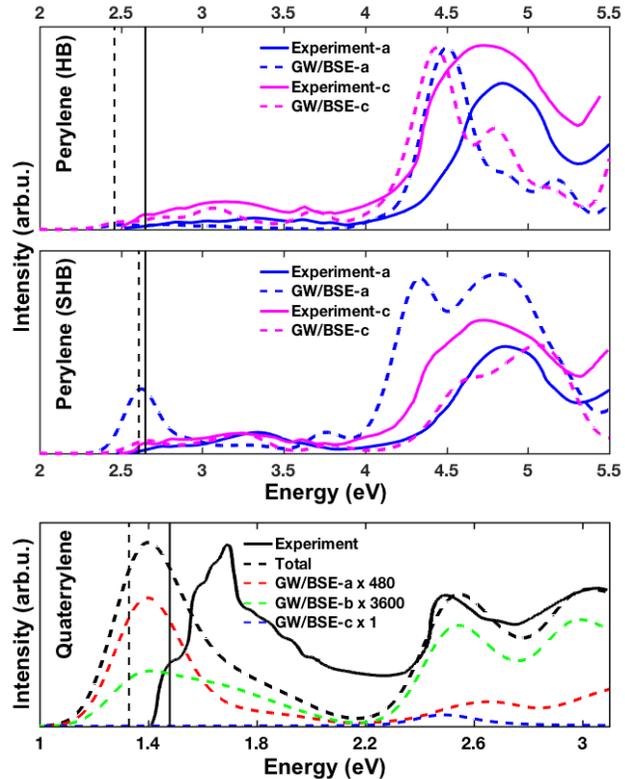

**Figure 4.** The experimental optical absorption spectra of two perylene polymorphs from Ref. 70, compared with $G_0W_0$+BSE@PBE absorption spectra along *a* (blue) and *c* (magenta). The experimental absorption spectrum of quaterrylene single crystal from Ref. 95 is compared with the total $G_0W_0$+BSE@PBE absorption spectrum (black), obtained by combining spectra calculated for light polarized along the three crystal axes with weights of 480:3600:1 along *a*, *b* and *c*, respectively. Experimental and computational optical gaps are shown in solid and dash vertical lines, respectively.

The energy conservation criterion ($E_S-2E_T$) is currently the primary descriptor thought to be associated with SF efficiency. The energy differences between the singlet exciton energy and twice the triplet exciton energy for perylene, quaterrylene, tetracene, pentacene, and rubrene are summarized in Table 1. $E_S$ and $E_T$ are compared with experimental data where available. As we have noted previously, $G_0W_0$+BSE@PBE systematically underestimates $E_S-2E_T$ (see also the above discussion of the limitations of $G_0W_0$+BSE).[49] This is particularly obvious for pentacene, for which SF is computationally predicted to be endoergic by 0.22 eV, inconsistent with experimental reports of fast SF with near 200% triplet yield.[42] Several recent benchmarks for molecular systems, using different codes, have shown that $G_0W_0$+BSE@PBE systematically underestimates both singlet and triplet excitation energies.[123,112-114] It is presently unknown whether the same trends persist in molecular solids, in part owing to lack of high-level reference data for periodic systems. From here on, we restrict the discussion to qualitative trends.

$E_S-2E_T$ values for both perylene polymorphs are much smaller than in quaterrylene, consistent with the trend reported for single molecules based on TDDFT calculations.[56] Based on the energy conservation criterion, the best SF candidates are pentacene and monoclinic rubrene,[49] followed closely by quaterrylene. $E_S-2E_T$ in quaterrylene is significantly higher than in tetracene and orthorhombic rubrene, where SF has been experimentally observed, therefore we conclude that SF is energetically favored with high efficiency in quaterrylene. Moreover, in a theoretical simulation of photovoltaic cells with different carrier multiplication absorbers, the highest solar conversion efficiency of 47.7% was predicted for an ideal two gap tandem photovoltaic device where the top cell is a SF absorber with a triplet gap of 0.84 eV.[15] Therefore, the triplet excitation energy of crystalline quaterrylene, 0.82 eV, which is also significantly smaller than all other materials in Table 1, makes it an ideal SF candidate for a tandem photovoltaic device.

**Effect of Crystal Packing on Exciton Wave-Functions**

While energy conservation is considered a necessary condition for SF to be thermodynamically favorable, the nature of the exciton wave-function may affect its efficiency. The degree of charge transfer character of the singlet exciton may determine the strength of its coupling with the multi-exciton state of two correlated triplet excitons localized on neighboring molecules, and thus influence SF dynamics.[7,8,9,20,22,138,139] The lowest energy singlet exciton wave-functions of perylene, quaterrylene, tetracene, pentacene, and rubrene are visualized in Figure 5 (plots of triplet exciton wave-functions are provided in the SI). The electron probability density is shown as yellow isosurfaces with respect to a fixed hole position indicated by a red dot. The eight crystals considered here exhibit different packing motifs, which give rise to different exciton spatial distributions. Hirshfeld surface analysis of the crystal packing is provided in the SI in order to relate the exciton wave-function distribution to the different intermolecular interactions in each system. Consistent trends in the singlet exciton distribution are found across different chromophores with similar crystal packing. Two of the systems considered here, perylene and rubrene, have polymorphic forms, in which the different crystal packing produces significantly different exciton wave-functions by modifying the intermolecular coupling.

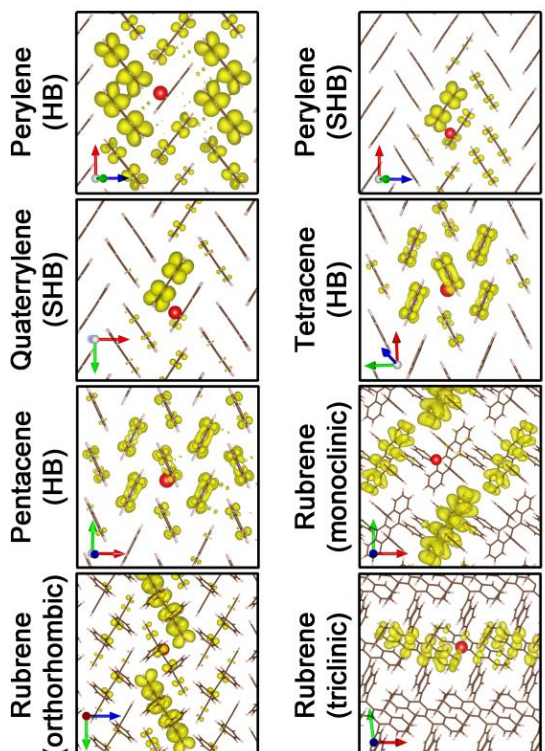

**Figure 5.** Singlet exciton wave functions for perylene, quaterrylene, tetracene, pentacene and rubrene crystals. Red dots indicate hole positions. Electron probability densities are shown in yellow.

The electron probability distribution is localized primarily on the molecules that have the strongest electronic coupling with the molecule on which the hole resides. In structures with strong cofacial π-stacking interactions, characterized by a high fraction of C⋯C intermolecular interactions (see SI), such as orthorhombic and triclinic rubrene, the electron and hole of the lowest energy singlet exciton typically reside on the cofacial neighbors. Based on double-Bader analysis, the combined electron probabilities for the two neighbors along the π-stacking direction in orthorhombic and triclinic rubrene are 75.7% and 80.1%, respectively. In the HB structure of β-perylene, tetracene and pentacene there are limited cofacial interactions, indicated by a smaller fraction of C···C intermolecular interactions and a large fraction of C···H interactions (see SI). As a result of every nearest neighbor in the molecular layer having approximately the same type and amount of intermolecular electronic coupling, the electron is distributed on several neighbors with approximately equal probabilities. When a crystal is comprised of HB-layers, as in β-perylene, tetracene and pentacene, the lowest-energy singlet exciton is typically delocalized within a single molecular layer[140,141] because the electronic coupling is strongest between molecules in the plane. As discussed in our previous work,[49] nearly no cofacial interactions exist between neighboring molecules in monoclinic rubrene. As a result, the electron is distributed on four neighboring molecules. The SHB structure of α-perylene and quaterrylene is comprised of slightly offset cofacially-stacked dimers arranged in a HB motif. As a consequence, their lowest-energy singlet excitons have features typical of both packing motifs. The significant cofacial overlap of the dimer pair, characterized by a high fraction of C···C interactions, results in 60.2% and 66.0% probabilities of the electron being found on the cofacial neighbor in perylene and quaterrylene, respectively. At the same time, the electron is dispersed over the molecular layer, characterized by a high fraction of C···H intermolecular interactions.

## A Two-Dimensional Descriptor for SF Efficiency

In Figure 6 the molecular crystals considered here are evaluated with respect to a two-dimensional descriptor for SF efficiency, based on the energy conservation criterion ($E_S-2E_T$) on the x-axis and the degree of singlet exciton charge transfer character (%CT) on the y-axis. A quantitative estimation of the degree of charge transfer character is provided by the double-Bader analysis method described above. The error bars represent the minimal and maximal %CT obtained for different hole positions, owing to slight changes in the electron probability distribution (see SI). Materials experimentally observed to exhibit SF are colored in red.

Figure 6 reveals trends across chemical families. For acenes and rylenes with the same crystal packing, namely tetracene and pentacene in the HB structure, and perylene and quaterrylene in the SHB structure, the larger molecules are more likely to exhibit

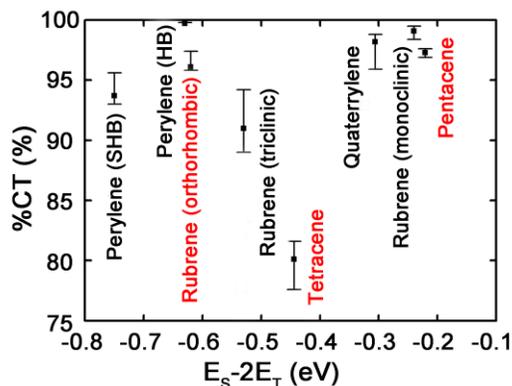

**Figure 6.** Crystalline perylene, quaterrylene, tetracene, pentacene and rubrene ranked with respect to a two-dimensional descriptor based on the energy conservation criterion, $E_S-2E_T$, and the degree of charge transfer character of the singlet exciton, %CT. Error bars indicate the minimal and maximal %CT obtained for different hole positions. Materials in which SF has been observed experimentally are shown in red. The most promising candidates are in the upper right corner.

SF, in terms of both the energy conservation criterion and the degree of CT (we note however, that even if this trend continued beyond pentacene, the larger acenes would not be practical for solar cell applications, owing to their instability). The trend for rylenes is consistent with the single molecule calculations of Ref. 56. This trend only holds within the same chemical family. For example, HB perylene is predicted to be less efficient than HB tetracene, although it is a larger molecule.

Of the chromophores shown in Figure 6, perylene has two polymorphs and rubrene has three polymorphs. In both cases, modifying the crystal structure leads to significant changes in the singlet and triplet excitation energies, as well as the character of the singlet exciton wave-function. For perylene, the HB structure, which has stronger intermolecular electronic coupling in the ground state, as indicated by its greater band dispersion, is ranked higher than the SHB structure with respect to both $E_S-2E_T$ and %CT. This is consistent with the conclusions of Refs. 20-22 that large electronic coupling leads to a high %CT and efficient SF in dimers and small aggregates of pentacene molecules in geometries derived from the HB crystal structure. The rubrene polymorphs, however, exhibit a different trend, whereby the monoclinic structure, which is weakly coupled in the ground state, as indicated by its flat bands, exhibits a high degree of CT character in the excited state.[49] Applying the methods of Refs. 20-22 to the rubrene polymorphs and additional chromophores other than pentacene in different packing arrangements than the HB structure may provide further insight. In summary, Figure 6 shows that intermolecular SF is the result of a complex interplay of the single molecule properties and the crystal packing. Therefore, predictive models should consider both.

The best candidates for intermolecular SF in the solid state, based on maximizing the two-dimensional descriptor ($E_S-2E_T$, %CT), are in the upper right corner of Figure 6. Pentacene, the quintessential SF material is ranked as the top candidate. The monoclinic form of rubrene is predicted to be close to pentacene in terms of both the energy conservation criterion and the degree of CT character of the singlet exciton,[49] and quaterrylene is a close third. Because SF has been observed experimentally in tetracene and orthorhombic rubrene, which are ranked low here due to low %CT and low $E_S-2E_T$, respectively, we expect SHB quaterrylene to exhibit SF. Based on the trend for the perylene polymorphs, growing quaterrylene in a HB structure could further enhance its SF performance. Compared to pentacene and monoclinic rubrene, quaterrylene has several additional advantages for device applications. Its narrow bandgap would enable absorbing a broader range of the solar spectrum, its triplet energy is in the optimal range for maximizing conversion efficiency in a tandem cell,[15] and finally, it is extremely stable, as it can be heated to over 500 °C without decomposing,[142] which would prevent degradation under operating conditions. Both perylene polymorphs have a high degree of CT character. However, they are ranked lower than tetracene and orthorhombic rubrene with respect to the energy conservation criterion. Therefore, they are unlikely to exhibit SF. They may be useful for harvesting sub-gap photons by TTA.[5,10,16,19]

## Conclusion

Inspired by the recent observation of SF in various rylene derivatives, we investigated the possibility of SF in crystalline perylene and quaterrylene. Many-body perturbation theory within the *GW*+BSE formalism was used to describe excited-state properties. A new method of double-Bader analysis was introduced for quantifying the degree of Frenkel vs. CT exciton character. A two-dimensional descriptor for SF efficiency was then proposed, based on maximizing the energy conservation criterion ($E_S-2E_T$) and the degree of CT character of the lowest energy singlet exciton (%CT). To assess their likelihood to exhibit SF, the two polymorphs perylene and quaterrylene were compared with other known and predicted SF materials, pentacene, tetracene, and the three polymorphs of rubrene.

Intermolecular SF in the solid state is the result of a complex interplay between the single molecule properties and the crystal structure. The comparison of chromophores of different chemical families in different crystal structures revealed trends across chemical families and elucidated the effect of crystal packing. Within the acene and rylene families, the larger chromophores are more likely to exhibit SF in terms of both the energy conservation criterion and the degree of singlet charge transfer character. For polymorphic chromophores the crystal structure significantly affects both $E_S-2E_T$ and %CT. In particular, the exciton wave-function is sensitive to the crystal packing with the electron probability distribution mainly localized on the molecules that have the strongest electronic coupling with the molecule on which the hole resides. This may enable the tuning of SF performance via crystal engineering.

Of the eight molecular crystals considered here, quaterrylene emerges as a promising candidate for the realization of solid-state SF-based solar cells. Based on $E_S-2E_T$ and %CT, quaterrylene is ranked as a close third after pentacene and monoclinic rubrene and could possibly be further enhanced by crystallization in a herringbone structure. From the device perspective, quaterrylene offers the further advantages of high stability, a narrow optical gap, and triplet excitation energy in the optimal range to maximize solar conversion efficiency in a tandem cell. Both polymorphs of perylene are excluded as SF candidates based on the energy conservation criterion. However, they are promising for increasing the conversion efficiency of solar cells by harvesting photons with energy below the optical gap via triplet-triplet annihilation.

**Acknowledgements**
Work at CMU was enabled by the generosity of the Charles E. Kaufman Foundation. Work at CPP was funded by National Science Foundation (NSF) Division of Materials Research through grant DMR-1637026. This research used resources of the Argonne Leadership Computing Facility (ALCF), which is a DOE Office of Science User Facility supported under Contract DE-AC02-06CH11357, and of the National Energy Research Scientific Computing Center (NERSC), a DOE Office of Science User Facility supported by the Office of Science of the U.S. Department of Energy under Contract No. DE-AC02-05CH11231.

**Supporting Information Available:** Comparison of the relaxed geometries of perylene and quaterrylene to experiment; *GW* calculations of perylene and quaterrylene molecules; *GW*+BSE results for tetracene; Detailed examples of the double-Bader exciton character analysis; Lowest energy triplet exciton wave functions; Hirshfeld surface analysis of crystal packing.

# References

(1) Kippelen, B.; Brédas, J.-L. Organic Photovoltaics. *Energy Environ. Sci.* **2009,** 2, 251–261.
(2) Clarke, T. M.; Durrant, J. R. Charge Photogeneration in Organic Solar Cells. *Chem. Rev.* **2010,** 110, 6736–6767.
(3) Heeger, A. J. 25th Anniversary Article: Bulk Heterojunction Solar Cells: Understanding the Mechanism of Operation. *Adv. Mater.* **2014,** 26, 10–28.
(4) Shockley, W.; Queisser, H. J. Detailed Balance Limit of Efficiency of p-n Junction Solar Cells. *J. Appl. Phys.* **1961,** 32, 510–519.
(5) Smith, M. B.; Michl, J. Singlet Fission. *Chem. Rev.* **2010,** 110, 6891–6936.
(6) Chan, W.-L.; Berkelbach, T. C.; Provorse, M. R.; Monahan, N. R.; Tritsch, J. R.; Hybertsen, M. S.; Reichman, D. R.; Gao, J.; Zhu, X.-Y. The Quantum Coherent Mechanism for Singlet Fission: Experiment and Theory. *Acc. Chem. Res.* **2013,** 46, 1321–1329.
(7) Monahan, N.; Zhu, X.-Y. Charge Transfer-Mediated Singlet Fission. *Annu. Rev. Phys. Chem.* **2015,** 66, 601–618.
(8) Berkelbach, T. C.; Hybertsen, M. S.; Reichman, D. R. Microscopic Theory of Singlet Exciton Fission. I. General Formulation. *J. Chem. Phys.* **2013,** 138, 114102.
(9) Berkelbach, T. C.; Hybertsen, M. S.; Reichman, D. R. Microscopic Theory of Singlet Exciton Fission. II. Application to Pentacene Dimers and the Role of Superexchange. *J. Chem. Phys.* **2013,** 138, 114103.
(10) Smith, M. B.; Michl, J. Recent Advances in Singlet Fission. *Annu. Rev. Phys. Chem.* **2013,** 64, 361–386.
(11) Lee, J.; Jadhav, P.; Reusswig, P. D.; Yost, S. R.; Thompson, N. J.; Congreve, D. N.; Hontz, E.; Van Voorhis, T. Baldo, M. A. Singlet Exciton Fission Photovoltaics. *Acc. Chem. Res.* **2013,** 46, 1300–1311.
(12) Bardeen, C. J. The Structure and Dynamics of Molecular Excitons. *Annu. Rev. Phys. Chem.* **2014,** 65, 127–148.
(13) Bardeen, C. J. Triplet Excitons: Bringing Dark States to Light. *Nat. Mater.* **2014,** 13, 1001–1003.
(14) Singh, S.; Jones, W. J.; Siebrand, W.; Stoicheff, B. P.; Schneider, W. G. Laser Generation of Excitons and Fluorescence in Anthracene Crystals. *J. Chem. Phys.* **1965,** 42, 330–342.
(15) Hanna, M. C.; Nozik, A. J. Solar Conversion Efficiency of Photovoltaic and Photoelectrolysis Cells with Carrier Multiplication Absorbers. *J. Appl. Phys.* **2006,** 100, 074510-1–074510-8.
(16) Paci, I.; Johnson, J. C.; Chen, X.; Rana, G.; Popović, D.; David, D. E.; Nozik, A. J.; Ratner, M. A.; Michl, J. Singlet Fission for Dye-Sensitized Solar Cells: Can a Suitable Sensitizer Be Found? *J. Am. Chem. Soc.* **2006,** 128, 16546–16553.
(17) Greyson, E. C.; Stepp, B. R.; Chen, X.; Schwerin, A. F.; Paci, I.; Smith, M. B.; Akdag, A.; Johnson, J. C.; Nozik, A. J.; Michl, J.; Ratner, M. A. Singlet Exciton Fission for Solar Cell Applications: Energy Aspects of Interchromophore Coupling. *J. Phys. Chem. B* **2010,** 114. 14223–14232.

(18) Chan, W. L.; Ligges, M.; Zhu, X.-Y. The Energy Barrier in Singlet Fission Can Be Overcome through Coherent Coupling and Entropic Gain. *Nat. Chem.* **2012,** 4, 840–845.

(19) Lin, Y. L.; Koch, M.; Brigeman, A. N.; Freeman, D. M. E.; Zhao, L.; Bronstein, H.; Giebink, N. C.; Scholes, G. D.; Rand, B. P. Enhanced Sub-Bandgap Efficiency of a Sollid-State Organic Intermediate Band Solar Cell Using Triplet-Triplet Annihilation. *Energy Environ. Sci.* **2017,** 10, 1465–1475.

(20) Beljonne, D.; Yamagata, H.; Brédas, J. L.; Spano, F. C.; Olivier, Y. Charge-Transfer Excitations Steer the Davydov Splitting and Mediate Singlet Exciton Fission in Pentacene. *Phys. Rev. Lett.* **2013,** 110, 226402-1–226402-5.

(21) Berkelbach, T. C.; Hybertsen, M. S.; Reichman, D. R. Microscopic Theory of Singlet Exciton Fission. III. Crystalline Pentacene. *J. Chem. Phys.* **2014,** 141, 074705-1–074705-12.

(22) Zeng, T.; Hoffmann, R.; Ananth, N. The Low-Lying Electronic States of Pentacene and Their Roles in Singlet Fission. *J. Am. Chem. Soc.* **2014,** 136, 5755–5764.

(23) Hummer, K.; Puschnig, P.; Ambrosch-Draxl, C. Lowest Optical Excitations in Molecular Crystals: Bound Excitons versus Free Electron-Hole Pairs in Anthracene. *Phys. Rev. Lett.* **2004,** 92, 147402-1–147402-4.

(24) Hummer, K.; Ambrosch-Draxl, C. Oligoacene Exciton Binding Energies: Their Dependence on Molecular Size. *Phys. Rev. B* **2005,** 71, 081202(R)-1–081202(R)-4.

(25) Ambrosch-Draxl, C.; Nabok, D.; Puschnig, P.; Meisenbichler, C. The Role of Polymorphism in Organic Thin Films: Oligoacenes Investigated from First Principles. *New J. Phys.* **2009,** 11, 125010.

(26) Zimmerman, P. M.; Bell, F.; Casanova, D.; Head-Gordon, M. Mechanism for Singlet Fission in Pentacene and Tetracene: from Single Exciton to Two Triplets. *J. Am. Chem. Soc.* **2011,** 133, 19944–19952.

(27) Schatschneider, B.; Monaco, S.; Tkatchenko, A.; Liang, J.-J. Understanding the Structure and Electronic Properties of Molecular Crystals Under Pressure: Application of Dispersion Corrected DFT to Oligoacenes. *J. Phys. Chem. A* **2013,** 117, 8323–8331.

(28) Rangel, T.; Berland, K.; Sharifzadeh, S.; Brown-Altvater, F.; Lee, K.; Hyldgaard, P.; Kronik, L.; Neaton, J. B. Structural and Excited-State Properties of Oligoacene Crystals from First Principles. *Phys. Rev. B* **2016,** 93, 115206-1–115206-16.

(29) Leng, X.; Feng, J.; Chen, T.; Liu, C.; Ma, Y. Optical Properties of Acene Molecules and Pentacene Crystal from the Many-body Green's Function Method. *Phys. Chem. Chem. Phys.* **2016,** 18, 30777–30784.

(30) Coto, P. B.; Sharifzadeh, S.; Neaton, J. B.; Thoss, M. Low-Lying Electronic Excited States of Pentacene Oligomers: A Comparative Electronic Structure Study in the Context of Singlet Fission. *J. Chem. Theory Comput.* **2015,** 11, 147-156.

(31) Sanders, S. N.; Kumarasamy, E.; Pun, A. B.; Steigerwald, M. L.; Sfeir, M. Y.; Campos, L. M. Singlet Fission in Polypentacene. *Chem* **2016,** 1, 505-511.

(32) Lee, J.; Bruzek, M. J.; Thompson, N. J.; Sfeir, M. Y.; Anthony, J. E.; Baldo, M. A. Singlet Exciton Fission in a Hexacene Derivative. *Adv. Mater.* **2013,** 25, 1445-1448.

(33) Thorley, K. J.; Finn, T. W.; Jarolimek, K.; Anthony, J. E.; Risko, C. Theory-Driven Insight into the Crystal Packing of Trialkylsilylethynyl Pentacenes. *Chem. Mater.* **2017,** 29, 2502-2512.

(34) Sutton, C.; Risko, C.; Brédas, J.-L. Noncovalent Intermolecular Interactions in Organic Electronic Materials: Implications for the Molecular Packing vs Electronic Properties of Acenes. *Chem. Mater.* **2016,** 28, 3-16.

(35) Sanders, S. N.; Kumarasamy, E.; Pun, A. B.; Trinh, M. T.; Choi, B.; Xia, J.; Taffet, E. J.; Low, J. Z.; Miller, J. R.; Roy, X.; Zhu, X.-Y.; Steigerwald, M. L.; Sfeir, M. Y.; Campos, L. M. Quantitative Intramolecular Singlet Fission in Bipentacenes. *J. Am. Chem. Soc.* **2015,** 137, 8965–8972.

(36) Korovina, N. V.; Das, S.; Nett, Z.; Feng, X.; Joy, J.; Haiges, R.; Krylov, A. I.; Bradforth, S. E.; Thompson, M. E. Singlet Fission in a Covalently Linked Cofacial Alkynyltetracene Dimer. *J. Am. Chem. Soc.* **2016,** 138, 617–627.

(37) Burdett, J. J.; Müller, A. M.; Gosztola, D.; Bardeen, C. J. Excited State Dynamics in Solid and Monomeric Tetracene: The Roles of Superradiance and Exciton Fission. *J. Chem. Phys.* **2010,** 133, 144506-1–144506-12.

(38) Burdett, J. J.; Bardeen, C. J. The Dynamics of Singlet Fission in Crystalline Tetracene and Covalent Analogs. *Acc. Chem. Res.* **2013,** 46, 1312–1320.

(39) Wilson, M. W. B.; Rao, A.; Johnson, K.; Gélinas, S.; Pietro, R. d.; Clark, J.; Friend, R. H. Temperature-Independent Singlet Exciton Fission in Tetracene. *J. Am. Chem. Soc.* **2013,** 135, 16680–16688.

(40) Thompson, N. J.; Wilson, M. W. B.; Congreve, D. N.; Brown, P. R.; Scherer, J. M.; Bischof, T. S.; Wu, M.; Geva, N.; Welborn, M.; Voorhis, T. V.; Bulović, V.; Bawendi, M. G.; Baldo, M. A. Energy Harvesting of Non-Emissive Triplet Excitons in Tetracene by Emissive PbS Nanocrystals. *Nat. Mater.* **2014,** 13, 1039–1043.

(41) Arias, D. H.; Ryerson, J. L.; Cook, J. D.; Damrauer, N. H.; Johnson, J. C. Polymorphism Influences Singlet Fission Rates in Tetracene Thin Films. *Chem. Sci.* **2016,** 7, 1185–1191.

(42) Rao, A; Wilson, M. W. B.; Hodgkiss, J. M.; Albert-Seifried, S.; Bässler, H.; Friend, R. H. Exciton Fission and Charge Generation via Triplet Excitons in Pentacene/C60 Bilayers. *J. Am. Chem. Soc.* **2010,** 132, 12698–12703.

(43) Sharifzadeh, S.; Darancet, P.; Kronik, L.; Neaton, J. B. Low-Energy Charge-Transfer Excitons in Organic Solids from First-Principles: The Case of Pentacene. *J. Phys. Chem. Lett.* **2013,** 4, 2197–2201.

(44) Tabachnyk, M.; Ehrler, B.; Gélinas, S.; Böhm, M. L.; Walker, B. J.; Musselman, K. P.; Greenham, N. C.; Friend, R. H.; Rao, A. Resonant Energy Transfer of Triplet Excitons from Pentacene to PbSe Nanocrystals. *Nat. Mater.* **2014,** 13, 1033–1038.

(45) Walker, B. J.; Musser, A. J.; Beljonne, D.; Friend, R. H. Singlet Exciton Fission in Solution. *Nat. Chem.* **2013,** 5, 1019–1024.

(46) Sharifzadeh, S.; Wong, C. Y.; Wu, H.; Cotts, B. L.; Kronik, L.; Ginsberg, N. S.; Neaton, J. B. Relating the Physical Structure and Optoelectronic Function of Crystalline TIPS-Pentacene. *Adv. Funct. Mater.* **2015,** 25, 2038–2046.

(47) Ma, L.; Zhang, K.; Kloc, C.; Sun, H.; Michel-Beyerle, M. E.; Gurzadyan, G. G. Singlet Fission in Rubrene Single Crystal: Direct Observation by Femtosecond Pump-Probe Spectroscopy. *Phys. Chem. Chem. Phys.* **2012,** 14, 8307–8312.

(48) Ryasnyanskiy, A.; Biaggio, I. Tripelt Exciton Dynamics in Rubrene Single Crystals. *Phys. Rev. B* **2011,** 84, 193203-1–193203-4.

(49) Wang, X.; Garcia, T.; Monaco, S.; Schatschneider, B.; Marom, N. Effect of Crystal Packing on the Excitonic Properties of Rubrene Polymorphs. *CrystEngComm* **2016,** 18, 7353–7362.

(50) Kerr, K. A.; Ashmore, J. P.; Speakman, J. C. The Crystal and molecular Structure of Quaterrylene: a Redetermination. *Proc. R. Soc. Lond. A* **1975,** 344, 199–215.

(51) Clar, E.; Schmidt, W. Correlations Between Photoelectron and Ultraviolet Absorption Spectra of Polycyclic Hydrocarbons. The Terrylene and Peropyrene Series. *Tetrahedron* **1978,** 34, 3219–3224.

(52) Rumi, M.; Zerbi, G. Vibrational and Nonlinear Optical Properties of Rylenes Calculated by ab initio Methods. *J. Chem. Phys*. **1998,** 108, 8662–8670.

(53) Malloci, G. Cappellini, G. Mulas, G. Mattoni, A. Electronic and Optical Properties of Families of Polycyclic Aromatic Hydrocarbons: A Systematic (Time-Dependent) Density Functional Theory Study. *Chem. Phys.* **2011,** 384, 19–27.

(54) Markiewicz, J. T.; Wudl, F. Perrylene, Oligorylenes, and Aza-Analogs. *ACS Appl. Mater. Interfaces* **2015,** 7, 28063–28085.

(55) Zhao, X.; Xiong, Y.; Ma, J.; Yuan, Z. Rylene and Rylene Diimides: Comparison of Theoretical and Experimental Results and Prediction for High-Rylene Derivatives. *J. Phys. Chem. A* **2016,** 120, 7554–7560.

(56) Minami, T.; Ito, S.; Nakano, M. Theoretical Study of Singlet Fission in Oligorylenes. *J. Phys. Chem. Lett.* **2012,** 3, 2719–2723.

(57) Miller, C. E.; Wasielewski, M. R.; Schatz, G. C. Modeling Singlet Fission in Rylene and Diketopyrrolopyrrole Derivatives: The Role of the Charge Transfer State in Superexchange and Excimer Formation. *J. Phys. Chem. C* **2017,** 121, 10345–10350.

(58) Takeda, Y.; Katoh, R.; Kobayashi, H.; Kotani, M. Fission and Fusion of Excitons in Perylene Crystal Studied with VUV and X-Ray Excitation. *J. Electron. Spectrosc. Relat. Phenom.* **1996,** 78, 423–426.

(59) Renaud, N.; Sherratt, P. A.; Ratner, M. A. Mapping the Relation between Stacking Geometries and Singlet Fission Yield in a Class of Organic Crystals. *J. Phys. Chem. Lett*. **2013,** 4, 1065–1069.

(60) Jiang, H.; Zhang, K. K.; Ye, J.; Wei, F.; Hu, P.; Guo, J.; Liang, C.; Chen, X.; Zhao, Y.; McNeil, L. E.; Hu, W.; Kloc, C. Atomically Flat, Large-Sized, Two-Dimensional Organic Nanocrystals. *Small* **2013,** 9, 990–995.

(61) Eaton, S. W.; Shoer, L. E.; Karlen, S. D.; Dyar, S. M.; Margulies, E. A.; Veldkamp, B. S.; Ramanan, C.; Hartzler, D. A.; Savikhin, S.; Marks, T. J.; Wasielewski, M. R. Singlet Exciton Fission in Polycrystalline Thin Films of a Slip-Stacked Perylenediimide. *J. Am. Chem. Soc.* **2013,** 135, 14701–14712.

(62) Mirjani, F.; Renaud, N.; Gorczak, N.; Grozema, F. C. Theoretical Investigation of Singlet Fission in Molecular Dimers: The Role of Charge Transfer States and Quantum Interference. *J. Phys. Chem. C* **2014,** 118, 14192–14199.

(63) Renaud, N.; Grozema, C. Intermolecular Vibrational Modes Speed Up Singlet Fission in Perylenediimide Crystals. *J. Chem. Phys. Lett.* **2015,** 6, 360–365.

(64) Le, A. K.; Bender, J. A.; Roberts, S. T. Slow Singlet Fission Observed in a Polycrystalline Perylenediimide Thin Film. *J. Phys. Chem. Lett.* **2016,** 7, 4922–4928.

(65) Würthner, F.; Saha-Möller, C. R.; Fimmel, B.; Ogi, S.; Leowanawat, P.; Schmidt, D. Perylene Bisimide Dye Assemblies as Archetype Functional Supramolecular Materials. *Chem. Rev.* **2016,** 116, 962–1052.

(66) Eaton, S. W.; Miller, S. A.; Margulies, E. A.; Shoer, L. E.; Schaller, R. D.; Wasielewski, M. R. Singlet Exciton Fission in Thin Films of tert-Butyl-Substituted Terrylenes. *J. Phys. Chem. A* **2015,** 119, 4151–4161.

(67) Margulies, E. A.; Miller, C. E.; Wu, Y.; Ma, L.; Schatz, G. C.; Young, R. M.; Wasielewski, M. R. Enabling Singlet Fission by Controlling Intramolecular Charge Transfer in π-stacked Covalent Terrylenediimide Dimers. *Nat. Chem.* **2016,** 8, 1120–1125.

(68) Margulies, E. A.; Logsdon, J. L.; Miller, C. E.; Ma, L.; Simonoff, E.; Young, R. M.; Schatz, G. C.; Wasielewski, M. R. Direct Observation of a Charge-Transfer State Preceding High-Yield Singlet Fission in Terrylenediimide Thin Films. *J. Am. Chem. Soc.* **2017,** 139, 663–671.

(69) Sutton, C.; Tummala, N. R.; Beljonne, D.; Brédas, J.-L. Singlet Fission in Rubrene Derivatives: Impact of Molecular Packing. *Chem. Mater.* **2017,** 29, 2777-2787.

(70) Tanaka, Jiro. The Electronic Spectra of Aromatic Molecular Crystals. II. The Crystal Structure and Spectra of Perylene. *Bull. Chem. Soc. Jpn.* **1963,** 36, 1237–1249.

(71) Ranganathan, A.; Kulkarni, G. U. Aromaticity in Benzene-like Rings – An Experimental Electron Density Investigation. *Proc. Indian Acad. Sci. (Chem. Sci.)* **2003,** 115, 637–647.

(72) Botoshansky, M.; Herbstein, F. H.; Kapon, M. Toward a Complete Description of a Polymorphic Crystal: The Example of Perrylene Redetermination of the Structures of the (Z = 2 and 4) Polymorphs. *Helv. Chim. Acta* **2003,** 86, 1113–1128.

(73) Smith, D. J.; Fryer, J. R. Molecular Detail in Electron Micrographs of Quaterrylene $C_{40}H_{20}$. *Nature* **1981,** 291, 481–482.

(74) Kobayashi, T.; Isoda, S. Lattice Images and Molecular Images of Organic Materials. *J. Mater. Chem.* **1993,** 3, 1–14.

(75) Schatschneider, B.; Monaco, S.; Liang, J.-J.; Tkatchenko, A. High-Throughput Investigation of the Geometry and Electronic Structures of Gas-Phase and Crystalline Polycyclic Aromatic Hydrocarbons. *J. Phys. Chem. C* **2014,** 118, 19964–19974.

(76) Hedin, L. New Method for Calculating the One-Particle Green's Function with Application to the Electron-Gas Problem. *Phys. Rev.* **1965,** 139, A796–A823.

(77) Hybertsen, M. S.; Louie, S. G. Electron Correlation in Semiconductors and Insulators: Band Gaps and Quasiparticle Energies. *Phys. Rev. B* **1986,** 34, 5390–5413.

(78) Rohlfing, M.; Louie, S. G. Electron-Hole Excitations and Optical Spectra from First Principles. *Phys. Rev. B* **2000,** 62, 4927–4944.

(79) Marom, N. Accurate Description of the Electronic Structure of Organic Semiconductors by *GW* Methods. *J. Phys.: Condens. Matter* **2017,** 29, 103003.
(80) Onida, G.; Reining, L.; Rubio, A. Electronic Excitations: Density-Functional versus Many-Body Green's-Function Approaches. *Rev. Mod. Phys.* **2002,** 74, 601–659.
(81) Tang, W.; Sanville, E.; Henkelman, G. A Grid-Based Bader Analysis Algorithm without Lattice Bias. *J. Phys.: Condens. Matter* **2009,** 21, 084204-1–084204-7.
(82) Yu, M.; Trinkle, D. R. Accurate and Efficient Algorithm for Bader Charge Integration. *J. Chem. Phys.* **2011,** 134, 064111-1–064111-8.
(83) Bader, R. F. Atoms in Molecules: A Quantum Theory. Oxford University Press, Oxford, UK, 1990.
(84) Holmes, D.; Kumaraswamy, S.; Matzger, A. J.; Vollhardt, K. P. C. On the Nature of Nonplanarity in the [N]Phenylenes. *Chem. Eur. J.* **1999,** 5, 3399-3412.
(85) Clark, S. J.; Segall, M. D.; Pickard, C. J.; Hasnip, P. J.; Probert, M. I. J.; Refson, K.; Payne, M. C. First Principles Methods Using CASTEP. *Z. Kristallogr. – Cryst. Mater.* **2005,** 220, 567–570.
(86) Perdew, J. P.; Burke, K.; Ernzerhof, M. Generalized Gradient Approximation Made Simple. *Phys. Rev. Lett.* **1996,** 77, 3865–3868.
(87) Perdew, J. P.; Burke, K.; Ernzerhof, M. Generalized Gradient Approximation Made Simple [Phys. Rev. Lett. 77, 3865 (1996)]. *Phys. Rev. Lett.* **1997,** 78, 1396–1396.
(88) Tkatchenko, A.; Scheffler, M. Accurate Molecular Van Der Waals Interactions from Ground-State Electron Density and Free-Atom Reference Data. *Phys. Rev. Lett.* **2009,** 102, 073005-1–073005-4.
(89) Deslippe, J.; Samsonidze, G.; Strubbe, D. A.; Jain, M.; Cohen, M. L.; Louie, S. G. BerkeleyGW: A Massively Parallel Computer Package for the Calculation of the Quasiparticle and Optical Properties of Materials and Nanostructures. *Comput. Phys. Commun.* **2012,** 183, 1269–1289.
(90) Giannozzi, P.; Baroni, S.; Bonini, N.; Calandra, M.; Car, R.; Cavazzoni, C.; Ceresoli, D.; Chiarotti, G. L.; Cococcioni, M.; Dabo, I.; Corso, A. D.; Gironcoli, S. d.; Fabris, S.; Fratesi, G.; Gebauer, R.; Gerstmann, U.; Gougoussis, C.; Kokalj, A.; Lazzeri, M.; Martin-Samos, L.; Marzari, N.; Mauri, F.; Mazzarello, R.; Paolini, S.; Pasquarello, A.; Paulatto, L.; Sbraccia, C.; Scandolo, S.; Sclauzero, G.; Seitsonen, A. P.; Smogunov, A.; Umari, P.; Wentzcovitch, R. M. QUANTUM ESPRESSO: A Molecular and Open-Source Software Project for Quantum Simulations of Materials. *J. Phys.: Condens. Matter* **2009,** 21, 395502-1–395502-19.
(91) Troullier, N.; Martins, J. L. Efficient Pseudopotentials for Plane-Wave Calculations. *Phys. Rev. B* **1991,** 43, 1993–2006.
(92) Fuchs, M.; Scheffler, M. Ab Initio Pseudopotentials for Electronic Structure Calculations of Poly-Atomic Systems Using Density-Functional Theory. *Comput. Phys. Commun.* **1999,** 119, 67–98.
(93) Deslippe, J.; Samsonidze, G.; Jain, M.; Cohen, M. L.; Louie, S. G. Coulomb-Hole Summations and Energies for *GW* Calculations with Limited Number of Empty Orbitals: A Modified Static Remainder Approach. *Phys. Rev. B* **2013,** 87, 165124-1–165124-6.
(94) Setyawan, W.; Curtarolo, S. High-Throughput Electronic Band Structure Calculations: Challenges and Tools. *Comput. Mater. Sci.* **2010,** 49, 299–312.

(95) Maruyama, Y.; Iwaki, T.; Kajiwara, T.; Shirotani, I.; Inokuchi, H. Molecular Orientation and Absorption Spectra of Quaterrylene Evaporated Film. *Bull. Chem. Soc. Jpn.* **1970,** 43, 1259–1261.
(96) Neaton, J. B.; Hybertsen, M. S.; Louie, S. G. Renormalization of Molecular Electronic Levels at Metal-Molecule Interfaces. *Phys. Rev. Lett.* **2006,** 97, 215405-1–215405-4.
(97) Thygesen, K. S.; Rubio, A. Renormalization of Molecular Quasiparticle Levels at Metal-Molecular Interfaces: Trends across Binding Regimes. *Phys. Rev. Lett.* **2009,** 102, 046802-1–046802-4.
(98) Freysoldt, C.; Rinke, P.; Scheffler, M. Controlling Polarization at Insulating Surfaces: Quasiparticle Calculations for Molecules Adsorbed on Insulator Films. *Phys. Rev. Lett.* **2009,** 103, 056803-1–056803-4.
(99) Lim, S.-H.; Bjorklund, T. G.; Spano, F. C.; Bardeen, C. J. Exciton Delocalization and Superradiance in Tetracene Thin Films and Nanoaggregates. *Phys. Rev. Lett.* **2004,** 92, 107402-1–107402-4.
(100) Bree, A.; Lyons, L. E. Photo- and Semi-Conductance of Organic Crystals. Part VI. Effect of Oxygen on the Surface Photo-Current and Some Photochemical Properties of Solid Anthracene. *J. Chem. Soc.* **1960,** 5206–5212.
(101) Hestand, N. J.; Yamagata, H.; Xu, B.; Sun, D.; Zhong, Y.; Harutyunyan, A. R.; Chen, G.; Dai, H.-L.; Rao, Y.; Spano, F. C. Polarized Absorption in Crystalline Pentacene: Theory and Experiment. *J. Phys. Chem. C* **2015,** 119, 22137–22147.
(102) Huang, L.; Liao, Q.; Shi, Q.; Fu, H.; Ma, J.; Yao, J. Rubrene Micro-Crystals from Solution Routes: Their Crystallography, Morphology and Optical Properties. *J. Mater. Chem.* **2010,** 20, 159–166.
(103) Irkhin, P.; Ryasnyanskiy, A.; Koehler, M.; Biaggio, I. Absorption and Photoluminescence Spectroscopy of Rubrene Single Crystals. *Phys. Rev. B* **2012,** 86, 085143-1–085143-13.
(104) Tavazzi, S.; Borghesi, A.; Papagni, A.; Spearman P.; Silvestri, L.; Yassar, A.; Camposeo, A.; Polo, M.; Pisignano, D. Optical Response and Emission Waveguiding in Rubrene Crystals. *Phys. Rev. B* **2007,** 75, 245416-1–245416-5.
(105) Mitrofanov, O.; Lang, D. V.; Kloc, C.; Wikberg, J. M.; Siegrist, T.; So, W.-Y.; Sergent, M. A.; Ramirez, A. P. Oxygen-Related Band Gap State in Single Crystal Rubrene. *Phys. Rev. Lett.* **2006,** 97, 166601-1–166601-4.
(106) Thorsmølle, V. K.; Averitt, R. D.; Demsar, J.; Smith, D. L.; Tretiak, S.; Martin, R. L.; Chi, X.; Crone, B. K.; Ramirez, A. P.; Taylor, A. J. Morphology Effectively Controls Singlet-Triplet Exciton Relaxation and Charge Transport in Organic Semiconductors. *Phys. Rev. Lett.* **2009,** 102, 017401-1–017401-4.
(107) Jundt, C.; Klein, G.; Sipp, B.; Moigne, J. Le; Joucla, M.; Villaeys, A. A. Exciton Dynamics in Pentacene Thin Films Studied by Pump-Probe Spectroscopy. *Chem. Phys. Lett.* **1995,** 241, 84–88.
(108) Tao, S.; Matsuzaki, H.; Uemura, H.; Yada, H.; Uemura, T.; Takeya, J.; Hasegawa, T.; Okamoto, H. Optical Pump-Probe Spectroscopy of Photocarriers in Rubrene Single Crystals. *Phys. Rev. B* **2011,** 83, 075204-1–075204-9.
(109) Chen, Y.; Lee, B.; Fu, D.; Podzorov, V. The Origin of a 650 nm Photoluminescence Band in Rubrene. *Adv. Mater.* **2011,** 23, 5370–5375.

(110) Herkstroeter, W. G.; Merkel, P. B. The Triplet State Energies of Rubrene and Diphenylisobenzofuran. *J. Photochem.* **1981,** 16, 331–341.

(111) Knight, J. W.; Wang, X.; Gallandi, L.; Dolgounitcheva, O.; Ren, X.; Ortiz, J. V.; Rinke, P.; Körzdörfer, T.; Marom, N. Accurate Ionization Potentials and Electron Affinities of Acceptor Molecules III: A Benchmark of *GW* Methods. *J. Chem. Theory Comput.* **2016,** 12, 615–626.

(112) Bruneval, F.; Hamed, S. M.; Neaton, J. B. A Systematic Benchmark of the Ab Initio Bethe-Salpeter Equation Approach for Low-Lying Optical Excitations of Small Organic Molecules. *J. Chem. Phys.* **2015,** 142, 244101-1–244101-10.

(113) Jacquemin, D.; Duchemin, I.; Blase, X. Benchmarking the Bethe-Salpeter Formalism on a Standard Organic Molecular Set. *J. Chem. Theory Comput.* **2015,** 11, 3290–3304.

(114) Rangel, T.; Hamed, S. M.; Bruneval, F.; Neaton, J. B. An Assessment of Low-Lying Excitation Energies and Triplet Instabilities of Organic Molecules with an Ab Initio Bethe-Salpeter Equation Approach and the Tamm-Dancoff Approximation. *J. Chem. Phys.* **2017,** 146, 194108-1–194108-8.

(115) van Setten, M. J.; Caruso, F.; Sharifzadeh, S.; Ren, X.; Scheffler, M.; Liu, F.; Lischner, J.; Lin, L.; Deslippe, J. R.; Louie, S. G.; Yang, C.; Weigend, F.; Neaton, J. B.; Evers, F.; Rinke, P. *GW*100: Benchmarking $G_0W_0$ for Molecular Systems. *J. Chem. Theory Comput.* **2015,** 11, 5665–5687.

(116) Jacquemin, D.; Duchemin, I.; Blondel, A.; Blase, X. Assessment of the Accuracy of the Bethe-Salpeter (BSE/*GW*) Oscillator Strengths. *J. Chem. Theory Comput.* **2016,** 12, 3969–3981.

(117) Jacquemin, D.; Duchemin, I.; Blondel, A.; Blase, X. Benchmark of Bethe-Salpeter for Triplet Excited-States. *J. Chem. Theory Comput.* **2017,** 13, 767–783.

(118) Blase, X.; Duchemin, I.; Jacquemin, D. The Bethe-Salpeter Equation in Chemistry: Relations with TD-DFT, Applications and Challenges. *Chem. Soc. Rev.* **2017,** DOI: 10.1039/c7cs00049a.

(119) Leng, X.; Yin, H.; Liang, D.; Ma, Y. Excitons and Davydov Splitting in Sexithiophene from First-Principles Many-Body Green's Function Theory. *J. Chem. Phys.* **2015,** 143, 114501-1–114501-8.

(120) Tkatchenko, A.; DiStasio, R. A.; Car, R.; Scheffler, M. Accurate and Efficient Method for Many-Body van der Waals Interactions. *Phys. Rev. Lett.* **2012,** 108, 236402-1–236402-5.

(121) Marom, N.; DiStasio, R. A.; Atalla, V.; Levchenko, S.; Reilly, A. M.; Chelikowsky, J. R.; Leiserowitz, L.; Tkatchenko, A. Many-Body Dispersion Interactions in Molecular Crystal Polymorphism. *Angew. Chem. Int. Ed.* **2013,** 52, 6629–6632.

(122) Curtis, F.; Wang, X.; Marom, N. Effect of Packing Motifs on the Energy Ranking and Electronic Properties of Putative Crystal Structures of Tricyano-1,4-dithiino[*c*]-isothiazole. *Acta Cryst. B* **2016,** 72, 562–570.

(123) Hung, L.; da Jornada, F. H.; Souto-Casares, J.; Chelikowsky, J. R.; Louie, S. G.; Öğüt, S. Excitation Spectra of Aromatic Molecules within a Real-Space GW-BSE Formalism: Role of Self-Consistency and Vertex Corrections. *Phys. Rev. B* **2016,** 94, 085125-1–085125-13.

(124) Kaplan, F.; Weigend, F.; Evers, F.; Setten, M. J. van Off-Diagonal Self-Energy Terms and Partially Self-Consistency in *GW* Calculations for Single Molecules: Efficient Implementation and Quantitative Effects on Ionization Potentials. *J. Chem. Theory Comput.* **2015,** 11, 5152–5160.

(125) Maebashi, H.; Takada, Y. Analysis of Exact Vertex Function for Improving on the GW Γ Scheme for First-Principles Calculation of Electron Self-Energy. *Phys. Rev. B* **2011,** 84, 245134-1–245134-13.

(126) Shishkin, M.; Marsman, M.; Kresse, G. Accurate Quasiparticle Spectra from Self-Consistent *GW* Calculations with Vertex Corrections. Phys. Rev. Lett. 2007, 99, 246403-1–246403-4.

(127) Ren, X.; Marom, N.; Caruso, F.; Scheffler, M.; Rinke, P. Beyond the *GW* Approximation: A Second-Order Screened Exchange Correction. Phys. Rev. B **2015,** 92, 081104(R)-1–081104(R)-6.

(128) Grüneis, A.; Kresse, G.; Hinuma, Y.; Oba, F. Ionization Potentials of Solids: The Importance of Vertex Corrections. Phys. Rev. Lett. **2014,** 112, 096401-1–096401-5.

(129) Gómez-Abal, R.; Li, X.; Scheffler, M. Influence of the Core-Valence Interaction and of the Pseudopotential Approximation on the Electron Self-Energy in Semiconductors. *Phys. Rev. Lett.* **2008,** 101, 106404-1–106404-4.

(130) Friedrich, C.; Schindlmayr, A.; Blügel, S. Elimination of the Linearization Error in *GW* Calculations Based on the Linearized Augmented-Plane-Wave Method. Phys. Rev. B **2006,** 74, 045104-1–045104-8.

(131) Bruneval, F.; Marques, M. A. L. Benchmarking the Starting Points of the *GW* Approximation for Molecules. *J. Chem. Theory Comput.* **2013,** 9, 324–329.

(132) Körbel, S.; Boulanger, P.; Duchemin, I.; Blase, X.; Marques, M. A. L.; Botti, S. Benchmark Many-Body *GW* and Bethe-Salpeter Calculations for Small Transition Metal Molecules. *J. Chem. Theory Comput.* **2014,** 10, 3934–3943.

(133) Marom, N.; Caruso, F.; Ren, X.; Hofmann, O. T.; Körzdörfer, T.; Chelikowsky, J. R.; Rubio, A.; Scheffler, M.; Rinke, P. Benchmark of *GW* Methods for Azabenzenes. *Phys. Rev. B* **2012,** 86, 245127-1–245127-16.

(134) Refaely-Abramson, S.; da Jornada, F. H.; Louie, S. G. Neaton, J. B. Origins of Singlet Fission in Solid Pentacene from an ab initio Green's-Function Approach. *arXiv.* **2017,** 1706.01564.

(135) Rohlfing, M.; Louie, S. G. Optical Excitations in Conjugated Polymers. *Phys. Rev. Lett.* **1999,** 82, 1959–1962.

(136) Peng, Q.; Niu, Y.; Wang, Z.; Jiang, Y.; Li, Y.; Liu, Y.; Shuai, Z. Theoretical Predictions of Red and Near-Infrared Strongly Emitting X-Annulated Rylenes. *J. Chem. Phys.* **2011,** 134, 074510-1–074510-10.

(137) Ohta, K.; Naitoh, Y.; Tominaga, K.; Hirota, N.; Yoshihara, K. Femtosecond Transient Absorption Studies of trans- and cis-1,3,5-Hexatriene in Solution. *J. Phys. Chem. A* **1998,** 102, 35–44.

(138) Yost, S. R.; Lee, J.; Wilson, M. W. B.; Wu, T.; McMahon, D. P.; Parkhurst, R. R.; Thompson, N. J.; Congreve, D. N.; Rao, A.; Johnson, K.; Sfeir, M. Y.; Bawendi, M. G.; Swager, T. M.; Friend, R. H.; Baldo, M. A.; Voorhis, T. V. A Transferable Model for Singlet-Fission Kinetics. *Nat. Chem.* **2014,** 6, 492–497.

(139) Pensack, R. D.; Tilley, A. J.; Parkin, S. R.; Lee, T. S.; Payne, M. M.; Gao, D.; Jahnke, A. A.; Oblinsky, D. G.; Li, P.-F.; Anthony, J. E.; Seferos, D. S.; Scholes, G. D. Exciton Delocalization Drives Rapid Singlet Fission in Nanoparticles of Acene Derivatives. *J. Am. Chem. Soc.* **2015,** 137, 6790–6803.

(140) Cudazzo, P.; Gatti, M.; Rubio, A. Excitons in Molecular Crystals from First-Principles Many-Body Perturbation Theory: Picene versus Pentacene. *Phys. Rev. B* **2012,** 86, 195307-1–195307-8.

(141) Cudazzo, P.; Sottile, F.; Rubio, A.; Gatti, M. Exciton Dispersion in Molecular Solids. *J. Phys.: Condens. Matter* **2015,** 27, 113204.

(142) Maruyama, Y.; Inokuchi, H.; Harada, Y. Electronic Properties of Quaterrylene, $C_{40}H_{20}$. *Bull. Chem. Soc. Jpn.* **1963,** 36, 1193–1198.